\documentclass[prb,superscriptaddress,twocolumn]{revtex4}
\usepackage{amsmath}
\usepackage{amsfonts}
\usepackage{amssymb}  
\usepackage[dvips]{graphicx}

\begin{document}
\bibliographystyle{apsrev}

\title{Titanium single electron transistor fabricated by electron-beam lithography}

\author{Mika A. Sillanp\"a\"a}
\email[]{Mika.Sillanpaa@iki.fi}
\affiliation{Low Temperature Laboratory, Helsinki University of
Technology P.O.Box 2200, FIN-02015 HUT, Finland}
\author{Pertti J. Hakonen}
\affiliation{Low Temperature Laboratory, Helsinki University of
Technology P.O.Box 2200, FIN-02015 HUT, Finland}

\begin{abstract}

A new method to fabricate non-superconducting mesoscopic tunnel junctions
by oxidation of Ti is presented.
The fabrication process uses conventional electron beam lithography and shadow
deposition through an organic resist mask.
Superconductivity in Ti is suppressed by performing the deposition
under a suitable background pressure.
We demonstrate the method by making a single electron transistor which operated at $T < 0.4$ K and
had a moderate charge noise of
$2.5 \times 10^{-3}$ e/$\sqrt{\mathrm{Hz}}$ at 10 Hz.
Based on nonlinearities in the current-voltage characteristics at higher voltages,
we deduce the oxide barrier height of approximately 110 mV.

\end{abstract}


\maketitle

\section{Introduction}

Single-charge devices have been studied intensively as possible alternatives for present day
electronics (see, e.g., Ref.\ \onlinecite{sets}).
Their operation is based on the fabrication of small enough tunnel junctions such that the
energy of charging the tunnel junction capacitor, $E_{C} = e^2 /2C$, where $e$ is the
electron charge and $C$ the junction capacitance, is high compared to thermal fluctuations.
Although direct applications of single electronics are still few, structures based on
mesoscopic tunnel junctions
have several fundamental uses in tunnel spectroscopy \cite{gueron,mina},
or as quantum limited detectors for the measurement of
charge quantum bits \cite{schon,korotkov1}.
In most of these applications it is beneficial, if not necessary, to have the spectroscopic
probe made from non-superconducting material, but simultaneously to maintain the essential parts of
the circuit superconducting in order to observe quantum coherent behavior.
Typically it is not possible to meet these criteria by
oxidation of the aluminum leads that are part of the sample.
In this work, we develop a reliable new method for fabrication
of normal-insulator-normal tunnel barriers, which is based on the existing principles of
e-beam lithography and shadow evaporation.

Electron-beam lithography with an organic polymer resist mask has
served as the straightforward, flexible method to fabricate Al tunnel junctions
\cite{offsetmask,fultondolan}.
Since Al oxidizes easily and forms a several nm thick native oxide layer, yielding an
almost vanishing tunnel conductance of mesoscopic
junctions, sufficiently thin Al tunnel barriers (1-2 nm) are made in relatively low
oxygen pressures of typically 0.1 mbar. Due to the high oxide barrier of Al$_{2}$O$_{3}$
of 1.5 V (Ref.\ \onlinecite{al_oxide}), current-voltage ($IV$) curves do not become nonlinear due to barrier
suppression effects at the low
voltages relevant to the single-charge or spectroscopic experiments.

The fact that Al, having the property of easy oxidation, is a superconductor
with bulk $T_{C}$ of about 1.2 K has certainly facilitated the way towards the nowadays
good understanding of mesoscopic charging effects and superconductivity. Nevertheless,
as already mentioned,
there are several needs for tunnel barriers having normal counter electrodes also in zero
external magnetic field. In fact, various techniques have been developed quite recently
for this particular purpose. Gu\'eron \textit{et al.} \cite{gueron} used thin, fully
oxidized Al layers of $2 \times 1.2$ nm on Cu. Some weakly oxidizing metals such as Ni and Gd
have been oxidized by reactive evaporation \cite{react_ni} or through oxygen glow discharge
\cite{gadolin}. Chromium forms a native oxide layer of 1-2 nm, with the
desired property that the metal
is not a superconductor. Normal-conducting SET transistors made of Cr have indeed been
demonstrated
\cite{kuzmin,scherer}, with controversial values for the barrier height
of the Cr oxide: 170 mV (Ref.\ \onlinecite{kuzmin}), or 740 mV (Ref.\ \onlinecite{scherer}).
However, controlled fabrication of Cr oxide junctions seems still
difficult. Our attempts to oxidize Cr in various pressures of pure oxygen, or in air,
consistently resulted in a short circuit in the junction.

\section{Fabrication techniques}
\label{sec:ti}

Titanium seems a promising candidate material for fabrication of tunnel barriers. It forms
a native oxide layer of $\simeq 1$ nm, consisting mainly
of TiO$_{2}$ (Ref.\ \onlinecite{vaquila}), which could be a suitable barrier as such.
Ti has been used to fabricate ultrasmall junctions
by anodic STM nano-oxidation \cite{stm_tit1,rt_set,stm_tit2}. These experiments
give somewhat inconsistent values for the oxide barrier height between 178 mV (Ref.\ \onlinecite{stm_tit2})
and 285 mV (Ref.\ \onlinecite{rt_set}).

Although bulk Ti is superconducting with a $T_{C}$ of 0.39 K, the transition temperature
decreases rapidly if the metal is dirty \cite{tit_tc}. Vacuum evaporated films of Ti,
even if deposited under UHV conditions, seem to be disordered enough that the $T_{C}$ falls
well below temperatures accessible by small dilution refrigerators. Our Ti films evaporated at 
$p = 1 \times 10^{-8}$ mbar exhibited a relatively large resistivity
$\rho = 1.8 \; \mu \Omega$m and a RRR of 1.02. According
to Ref.\ \onlinecite{tit_tc}, a $T_{C}$ of 100 mK corresponds to a resistivity $\simeq 0.4$ $\mu \Omega$m.
Since our films were considerably dirtier, we expect them to stay normal over the
accessible temperature range down to 100 mK.

We used a standard double layer resist mask composed of 480 nm of copolymer and 120 nm of PMMA
for the Ti shadow evaporation. Ti was e-beam deposited at room temperature at a rate of 1 nm/s
at $p = 1 \times 10^{-8}$ mbar.
The deposited films seemed to have a tendency to spread some 50 nm around the
main line. This behavior was markedly pronounced in films evaporated at higher pressures of
$5 \times 10^{-6}$ mbar, where the metal had spread over the whole undercut
area even several $\mu$m wide. Since the spreading was strongly pressure dependent,
we believe it was caused by scattering of Ti atoms in the vacuum region between the surface
of the resist and the substrate. We first attempted to make tunnel junctions by e-beam depositing
a second Ti layer after oxidation of the first layer. This always resulted in a short circuit,
independent of the way the oxidation was performed. The problem was solved by making
the counter electrodes by thermal
evaporation of a Cu layer on top of the oxidized Ti. Apparently the oxide was
broken during the second e-beam deposition, possibly by x-rays coming from the
deposition target.

Along these lines, SET transistors were fabricated by first
patterning the island by e-beam evaporation of 15 nm of Ti, followed by oxidation
for 45 min in air. The electrodes were then made by thermal evaporation of 25 nm of Cu. 
We made a half dozen SET samples, all of which had reasonable room temperature
resistances, scattered between 50 k$\Omega$ and 15 M$\Omega$. One sample was cooled down
to 100 mK. Initially it had a tunnel resistance $R_{T} \simeq 50$ k$\Omega$, but the
resistance had increased during storage before cooldown
up to $\sim 100$ M$\Omega$. This tendency of resistance increase with
time seemed to be characteristic for all the fabricated junctions.

\section{Characterization of T\lowercase{i} SET}

\begin{figure}[htb]
\center
\includegraphics[width=0.95\linewidth]{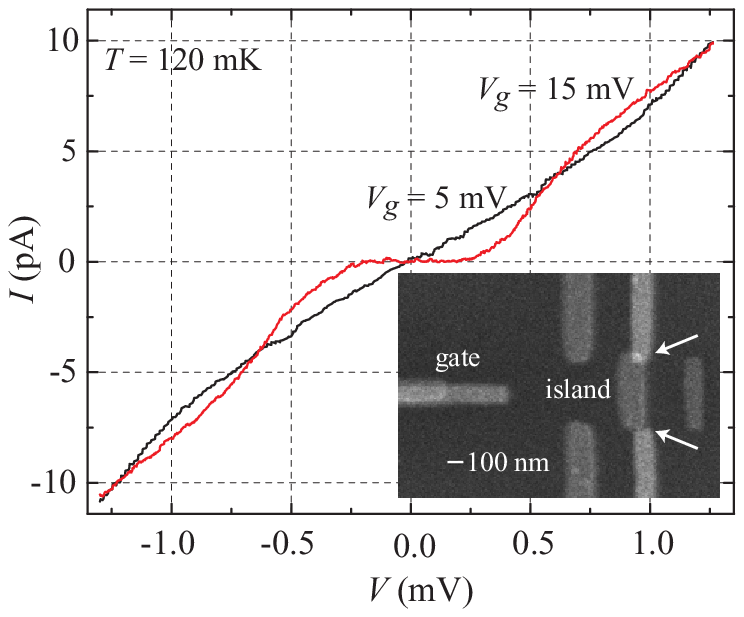}
\caption{$IV$-characteristics of Ti SET in the Coulomb blockade regime,
with two extreme gate voltages giving the largest and the smallest
width of the Coulomb gap.
Inset: image of a sample similar to the measured one. The tunnel
junctions are marked by arrows.}
\label{fig:iv}
\end{figure}

The measured SET device was rather asymmetrical (see inset of Fig.\ \ref{fig:iv}). The
bigger junction had a geometrical area of about $100 \times 100$ nm$^{2}$, but
in the smaller junction the films were hardly overlapping at all,
and thus it was difficult to deduce its area accurately.
The sample was cooled down to 100 mK in a plastic dilution fridge, the leads
being filtered with 0.7 m of Thermocoax cable. The sample was characterized using
voltage bias. $IV$-characteristics recorded at
$T = 120$ mK are shown in Fig.\ \ref{fig:iv}. Due to the high total tunnel
resistance of the two junctions, $R_{T} \simeq 100$ M$\Omega$, current noise
in the measurement system was a problem. The curves shown in Fig.\ \ref{fig:iv}
are a result of averaging over about 20 minutes. Gate modulation of the transistor
was visible below 0.4 K. Immediately after cooldown
there were strong fluctuations of the background charge, and the $IV$-curves didn't
stay stable. After several days, however, the background charge fluctuations
settled down and the gate operation point did not change
noticeably over tens of hours.

From the maximum width of the Coulomb blockade current plateau around zero voltage,
$V_{C} = C e \simeq 0.3$ mV, we derive the total island capacitance $C \simeq 0.53$ fF,
and a charging energy of 1.8 K.

\begin{figure}[htb]
\center
\includegraphics[width=0.95\linewidth]{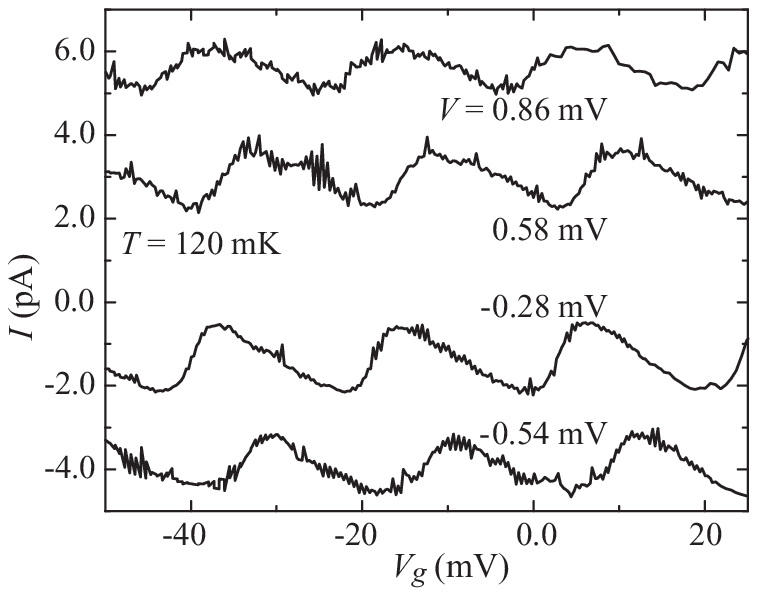}
\caption{Gate modulation curves of the transistor recorded at bias
voltages indicated in the figure.}
\label{fig:gate}
\end{figure}

Gate modulation curves of the
transistor at $T = 120$ mK are shown in Fig.\ \ref{fig:gate}.
Due to asymmetry of the device, the curves have different slopes at different
sides of the maxima.
Just beyond the Coulomb blockade threshold,
absolute values of the positive and
negative slope of the current modulation, at a fixed bias voltage, are given by

\begin{equation}
dI/dV_{g} = e/(R_{1,2} C \Delta V_{g}),
\label{eq:slope}
\end{equation}
where $\Delta V_{g}$ is the gate period.
In an asymmetrical device, the slopes are different because $R_{1} \ne R_{2}$.
On the basis of Eq.\ \ref{eq:slope}, we derive $R_{1} = 31$ M$\Omega$,
$R_{2} = 115$ M$\Omega$, and $R_{1}/R_{2} = 3.7$. We thus would expect
a total tunnel resistance of $R_{T} \simeq 146$ M$\Omega$,
which is of the same order as the value $R_{T} \simeq 100$ M$\Omega$
estimated directly from the $IV$-characteristics at low voltages \cite{rt_approx}.

Assuming the capacitance is dominated
by the contribution from the larger junction having the area
$A = (100$ nm$)^{2}$, and using the oxide thickness $d \simeq 16$
\AA \ determined in section \ref{sec:highv}, and the plate
capacitor formula $C = \epsilon \epsilon_{0}A/d$, we get a
dielectric constant $\epsilon \simeq 7.6$.
This value is drastically smaller than the
dielectric constants reported in literature for thin films of TiO$_{2}$,
$\epsilon \simeq 40 - 90$ (Refs.\ \onlinecite{perm1,perm2,perm3}).
In fact, such a high permittivity would make the capacitance so large that
single-charge effects would not be observable under our experimental conditions.
Reduced dielectric constants have been reported in tunnel barriers
made from other materials as well \cite{al_oxide,scherer}.

As discussed in section \ref{sec:ti}, we do not expect our Ti films
to go superconducting in the accessible temperature range. In a SET having
a superconducting island but normal electrodes, 
the Coulomb blockade plateau is widened at both sides by the amount $eV = 2 \Delta$
due to the energy gap $\Delta$ of the superconductor. Thus, when an
external magnetic field is applied to suppress superconductivity, onset
of current should switch approximately by the voltage $V = 2 \Delta /e$.
We made this test by recording the current at a fixed voltage 0.32 mV
just beyond the onset of the
current branch in a zero field, and in a high enough applied field $B = 150$ mT.
Results of these two measurement series averaged to the same with standard errors
of mean $\delta_{I} \simeq 15$ fA. We assume the effect of
superconductivity immediately beyond the onset of current can
be approximated simply by shifting the $IV$-curve along the voltage axis.
$\delta_{I}$ then corresponds to a shift in voltage by
$\delta_{V} \simeq 2 \mu$V, which determines the
minimum detectable gap $\Delta_{m} = \delta_{V}/2 \simeq 1 \mu$V.
Thus we have a minimum detectable
critical temperature $T_{C,m} = \Delta_{m} / 1.764 k_{B} \simeq 10$ mK.
When $T_{C,m}$ is compared to the measurement temperature of 100 mK,
we conclude that the Ti films were not superconducting at 100 mK, in
agreement with Ref.\ \onlinecite{tit_tc}.

\begin{figure}[htb]
\center
\includegraphics[width=0.92\linewidth]{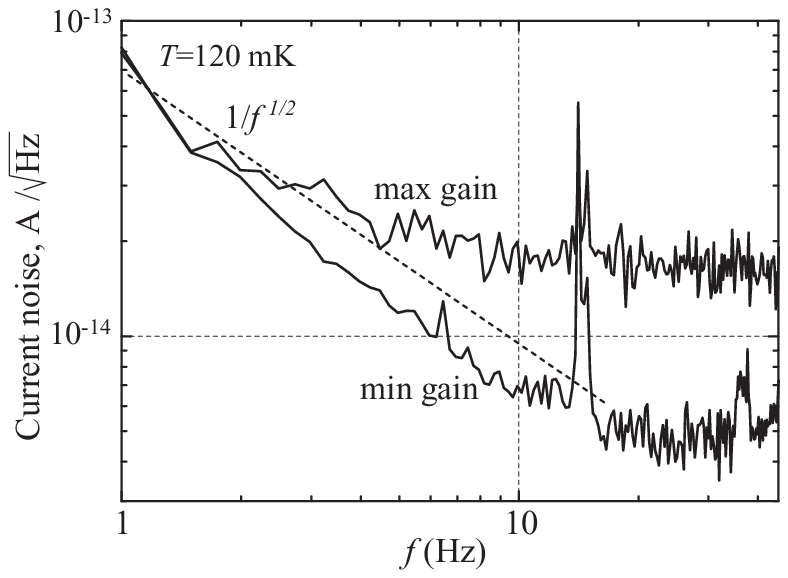}
\caption{Current noise amplitude of our Ti SET at bias voltage $V = -0.54$ mV,
for the maximum gain 6.5 pA/e, and for the minimum gain. At 10 Hz,
the current noise is 16 fA/$\sqrt{Hz}$, which corresponds to input charge noise of
$2.5 \times 10^{-3}$ e/$\sqrt{Hz}$. The dashed line depicts a $1/f$
behavior.}
\label{fig:noise}
\end{figure}

Current noise in the output of the Ti SET was measured in
the voltage-biased configuration at different
gate operation points of the transistor, and at different bias voltages. The noise spectra
shown in Fig.\ \ref{fig:noise} were measured at the bias $V = - 0.54$ mV
(lowest curve in Fig.\ \ref{fig:gate}), in the region
of maximum current modulation. Clearly, the noise at the maximum
gain is larger than at the minimum gain.
Below 10 Hz, the noise power follows roughly a $1/f$ behavior
expected from a collection of two-level fluctuators.
At higher bias points, the noise saturates at values comparable to
those of maximum gain at low bias.
At 10 Hz at the maximum gain of $g \simeq 6.5$ pA/e, we have a
current noise $I_{N} \simeq 16$ fA/$\sqrt{Hz}$,
which is translated to equivalent input charge noise $I_{Q} = I_{N}/g
\simeq 2.5 \times 10^{-3}$ e/$\sqrt{Hz}$. This figure is similar
or slightly higher than what we have observed in ordinary Al SETs
or what has been reported \cite{mooij_jap,noise_gain}, but considerably higher
than the lowest figures so far, $8 \times 10^{-6}$ e/$\sqrt{Hz}$,
observed in a stacked design \cite{stacknoise}.

\section{High voltage behavior}
\label{sec:highv}

At voltages comparable to the height of the potential barrier in the
insulating oxide, $IV$-curves become
nonlinear as a consequence of suppression of the barrier. This opens up the possibility to
characterize the barrier in terms of its height $\varphi$ and width $\Delta d$.
Several authors \cite{kuzmin,scherer} have used a formula derived
by Simmons \cite{simmons1} for the tunnel current density as


\begin{equation}
\begin{split}
j = j_{0} \{ & \bar{\varphi} \exp (-A \bar{\varphi}^{1/2} ) -\\&
(\bar{\varphi} + eV) \exp (- A (\bar{\varphi} + eV)^{1/2} ) \},
\end{split}
\label{eq:simmons}
\end{equation}
where $j_{0} = e/(2 \pi h \Delta d^{2} \beta^{2})$, $A = 4 \pi \beta \Delta d (2m)^{1/2}/h$,
and $\beta \approx 1$ is a correction factor discussed below in more detail.
The expression holds for a barrier of arbitrary shape and voltage provided 
$\bar{\varphi}$ is the mean barrier height above the Fermi level of the
negatively biased electrode, and $\Delta d$ is the width of the barrier at the
same position. For a symmetrical, rectangular barrier at low voltages
$eV < \varphi$, $\bar{\varphi} = \varphi - eV/2$,
and $\Delta d$ is constant and equal to the actual barrier width, $d$.

Large scale $IV$-characteristics of our Ti SET are shown in Fig.\ \ref{fig:large_iv}.
The asymmetry which is dependent on current direction can be explained
by a tunnel barrier which is asymmetrical due to different work functions
of the electrodes. In fact, our situation is rather complicated due to the presence
of two asymmetrical tunnel junctions having different resistances. However,
because the ratio of junction resistances was of the order 4, we made the
simplifying assumption of ignoring the smaller resistance junction in
fitting the Simmons' theory to our data.

As shown in the discussion of asymmetrical tunnel barriers in Ref.\ \onlinecite{solids},
the differential conductance of a barrier between electrodes having different
work functions $\Phi_{1}$ and $\Phi_{2}$ is symmetric about
the voltage $eV = \Phi_{1} - \Phi_{2} = \varphi_{1} - \varphi_{2}$.
Indeed, our data has this symmetry property
with respect to a +15 mV shift in conductance if the Coulomb blockade contribution
is excluded (data not shown). At larger
voltages, there is a similar symmetry about -45 mV. These shifts can be traced
back to the larger and smaller resistance junctions, respectively. Ratio of
the shifts agrees reasonably well with the estimated junction resistance ratio.
These data thus indicate a work function difference $\Delta \Phi$ $(=\Delta \varphi)$ of 10-15 mV
for the Cu-Ti pair in this configuration \cite{work_f}.


Although the correction factor $\beta \approx 1$ has been ignored in earlier
studies on symmetric junctions \cite{kuzmin,scherer}, it should not be neglected
since otherwise an error of tens of \% in the fitted parameters is
expected. This cast some doubts on the parameters obtained
for Cr junctions in Refs.\ \onlinecite{kuzmin,scherer}. In our case it is
necessary to include $\beta$, since it is the only factor
that gives rise to asymmetry in the $IV$-characteristics \cite{hartman}.

Let us now consider the asymmetrical, trapezoidal barrier with
heights $\varphi_{1}$ and $\varphi_{2}$
(Refs.\ \onlinecite{hartman,simmons2}), shown in the inset of Fig.\ \ref{fig:large_iv} (a).
If electrode 1 is negatively biased, the net electrical current density flows
to direction 1 and is denoted by $j_{1}$. In the case of opposite polarity,
the net current density flows to direction 2 and is denoted by $j_{2}$.
For the asymmetrical barrier, generally $j_{1} \ne j_{2}$. In the following,
the subscripts 1 and 2 refer to $j_{1}$ and $j_{2}$, respectively.

\subsection{Simmons' theory, intermediate voltages: $eV < \varphi_{1}$}

The average barrier heights at voltages satisfying
$eV < \varphi_{1}$ (without loss of generality,
we take $\varphi_{1}$ to be the lower barrier) are equal and are given by
$\bar{\varphi_{1}}(V) = \bar{\varphi_{2}}(V) =(\varphi_{1} + \varphi_{2} -eV)/2$.
The net current density $j_{1}$ is given by Eq.\ \ref{eq:simmons}, with

\begin{equation}
\beta_{1} = 1-(\varphi_{2}-\varphi_{1}-eV)^2/ [ 24(\varphi_{1}+\varphi_{2}-eV)^2 ].
\label{eq:beta1}
\end{equation}
The net current density $j_{2}$ is given by Eq.\ \ref{eq:simmons} as well, with

\begin{equation}
\beta_{2} = 1-(\varphi_{2}-\varphi_{1}+eV)^2/ [24 (\varphi_{1}+\varphi_{2}-eV)^2 ].
\label{eq:beta2}
\end{equation}
Equation \ref{eq:simmons}, when substituted with Eqs.\ \ref{eq:beta1} and \ref{eq:beta2}, should
describe the $IV$-characteristics at intermediate voltages $eV < \varphi_{1}$.
Because the area of the
large resistance junction was not known accurately, we scaled the current density
with a free fitting parameter. Before fitting, the voltages were multiplied
by the factor $(1+R_{2}/R_{1})^{-1} \simeq 0.79$ to account for division
of the applied voltage across the two junctions, using
their zero-voltage
resistance ratio $R_{1}/R_{2} \approx 3.7$. However, the fit turned out to be poor
especially at low voltages, in accord with
experiments on Cr junctions \cite{kuzmin,scherer}. In contrast to
the experimental data, theory predicts that at low voltages $j_{1}$ and $j_{2}$
are almost equal.
The discrepancy could not likely be explained by effect of the second
junction either.
Fit of the whole experimental data to $j_{1}$ and $j_{2}$ determined the parameters
$\varphi_{1} \simeq 120$ mV, $\varphi_{2} \simeq 140$ mV, and $d \simeq 16$ \AA.
However, the reliability of these fits is weakened
by the fact that several other parameter combinations resulted
in seemingly similar curves.
Barrier heights in this range are, however, favored by
other arguments as discussed in the next paragraphs.

\subsection{Simmons' theory, high voltages: $eV > \varphi_{1}$}

\begin{figure}[htb]
\center
\includegraphics[width=0.95\linewidth]{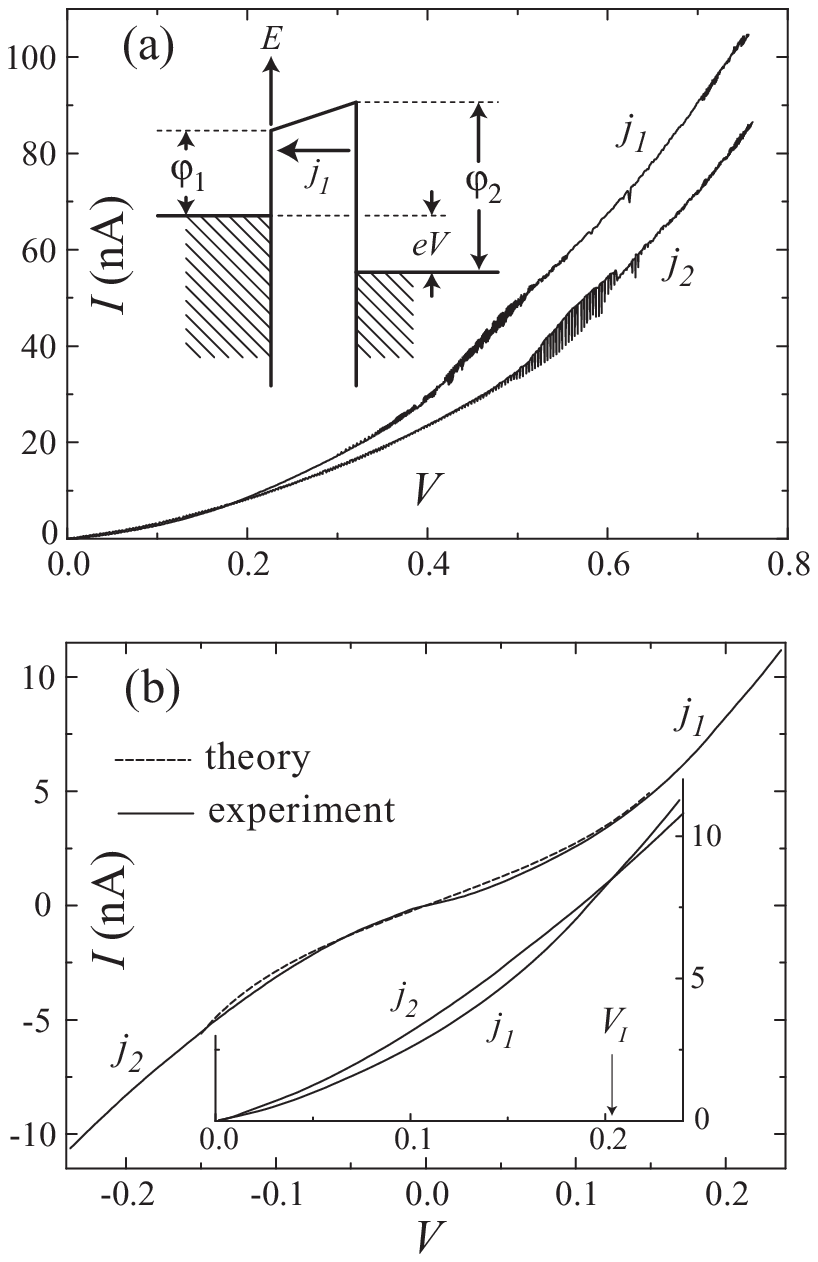}
\caption{(a) High-voltage $IV$-curves of the Ti SET, measured at $T = 120$ mK.
The positive and negative voltage parts of the $IV$-curve $j_{1}$ and $j_{2}$ are drawn
on the same axes. In the inset is shown the electron potential energy diagram for
an asymmetrical barrier biased in the
intermediate voltage range such that the net electrical current is in the direction 1.
Panel (b) shows the fit to Simmons' theory,
inset: $j_{1}$ and $j_{2}$ drawn on the same axes, with their intersection
voltage marked as $V_{I}$.}
\label{fig:large_iv}
\end{figure}

The intersection voltage of the two current branches $j_{1}$ and $j_{2}$,
marked as $V_{I}$ in the data in Fig.\ \ref{fig:large_iv}, offers another,
probably better, way to determine
the barrier height. The intersection occurs in the high voltage regime,
$eV > \varphi_{1}$. The currents are still given by Eq.\ \ref{eq:simmons},
but now with $\bar{\varphi} = \varphi_{1}/2$,
$\Delta d = d \varphi_{1}/(\varphi_{1}-\varphi_{2} + eV)$ for $j_{1}$,
and $\bar{\varphi} = \varphi_{2}/2$ and 
$\Delta d = d \varphi_{2}/(\varphi_{2}-\varphi_{1} + eV)$ for $j_{2}$.
For both $j_{1}$ and $j_{2}$, $\beta \simeq 23/24$.

Numerical calculations of the intersection voltage
indicate $V_{I} \simeq 1.5 \varphi_{1}$ if $\varphi_{1}$ and
$\varphi_{2}$ differ by at most 10-20 \%. From Fig.\ \ref{fig:large_iv}
we have $V_{I} \simeq 205$ mV. Thus, after correcting for voltage
division due to the second junction, we have $\varphi_{1} \simeq 110$ mV.
$\varphi_{2}$ is calculated with the independently determined value
$\Delta \varphi \simeq 15$ mV, to determine $\varphi_{2} \simeq 125$ mV.
These parameter values are in reasonable agreement with the fits to the
Simmons' theory in the intermediate-voltage regime, but not with
results of former work: 178 mV (Ref.\ \onlinecite{stm_tit2}),
285~mV (Ref.\ \onlinecite{rt_set}).

Around voltages of approximately +0.4 V and -0.5 V, there is a noisy section in the
$IV$-curves, followed by a change in resistance. Similar effects were
observed in Cr oxide junctions by Scherer \cite{scherer}, and interpreted
as being caused by charge traps in the barriers activating by a suitably high
electric field, thus giving rise to additional current paths. The
enhanced conductance in our case just beyond the noisy section agrees well
with this explanation.


\section{Summary and conclusions}



We have developed a straightforward and reliable way to fabricate
normal tunnel barriers by oxidation of titanium in air. The method
can find several uses in fundamental research or in applications. We use
the method to fabricate a single electron transistor which performs
comparably to traditional Al SETs.
Nonlinearities in the $IV$-characteristics at higher voltages indicate
the height of the TiO$_{2}$ barrier of 110 mV, and width of 16 \AA.

\begin{acknowledgments}

The authors would like to thank Hansj\"org Scherer for useful information.
This research was supported in part by Emil Aaltonen foundation,
and by the Human Capital and Mobility Program ULTI of the European Union.

\end{acknowledgments}


\begin{thebibliography}{99}
\bibitem{rt_approx}It is rather difficult to deduce $R_{T}$ from the $IV$-curve.
At low voltages the curve has nonlinearities due to Coulomb effects, and at higher
voltages, due to barrier suppression.
\end{thebibliography}
\end{document}